\documentclass[useAMS, usenatbib]{mn2e}
\usepackage{amsmath}
\usepackage{graphicx}

\def\be{\begin{equation}}
\def\ee{\end{equation}}
\def\ba{\begin{eqnarray}}
\def\ea{\end{eqnarray}}
\def\sl{\mathrm{sl}}
\def\lb{\mathrm{lb}}

\newcommand{\pl}{\Omega_{\rm{pl}}}
\newcommand{\ps}{\Omega_{\rm{ps}}}

\newcommand{\thetaSL}{\theta_{\rm{sl}}}
\newcommand{\thetaSB}{\theta_{\rm{sb}}}
\newcommand{\thetaLB}{\theta_{\rm{lb}}}

\newcommand{\hatS}{\hat{{\bf S}}}
\newcommand{\hatL}{\hat{{\bf L}}}
\newcommand{\hatLb}{\hat{{\bf L}}_b}

\def\go{\mathrel{\raise.3ex\hbox{$>$}\mkern-14mu
             \lower0.6ex\hbox{$\sim$}}}

\def\lo{\mathrel{\raise.3ex\hbox{$<$}\mkern-14mu
             \lower0.6ex\hbox{$\sim$}}}

\begin{document}
\title[Chaotic Dynamics of Stellar Spin]
{Chaotic Dynamics of Stellar Spin Driven by Planets Undergoing 
Lidov-Kozai Oscillations: Resonances and Origin of Chaos}
\author[Natalia I Storch and Dong Lai]
{Natalia I Storch\thanks{Email: nis22@cornell.edu, dong@astro.cornell.edu}
and Dong Lai\\
Center for Space Research, Department of Astronomy, Cornell University, Ithaca,
NY 14853, USA\\}

\pagerange{\pageref{firstpage}--\pageref{lastpage}} \pubyear{2014}

\label{firstpage}
\maketitle

\begin{abstract}
Many exoplanetary systems containing hot Jupiters are found to possess
significant misalignment between the spin axis of the host star and
the planet's orbital angular momentum axis.  A possible channel for
producing such misaligned hot Jupiters involves Lidov-Kozai
oscillations of the planet's orbital eccentricity and inclination
driven by a distant binary companion. In a recent work (Storch,
Anderson \& Lai 2014), we have shown that a proto-hot Jupiter
undergoing Lidov-Kozai oscillations can induce complex, and often
chaotic, evolution of the spin axis of its host star.  Here we explore
the origin of the chaotic spin behavior and its various features in an
idealized non-dissipative system where the secular oscillations of the
planet's orbit are strictly periodic.  Using Hamiltonian perturbation
theory, we identify a set of secular spin-orbit resonances in the
system, and show that overlaps of these resonances are responsible for
the onset of wide-spread chaos in the evolution of stellar spin axis.
The degree of chaos in the system depends on the adiabaticity parameter
$\epsilon$, proportional to the ratio of the Lidov-Kozai nodal precession
rate and the stellar spin precession rate, and thus depends on the
planet mass, semi-major axis and the stellar rotation rate.  For
systems with zero initial spin-orbit misalignment, our theory explains
the occurrence (as a function of $\epsilon$) of ``periodic islands'' in
the middle of a ``chaotic ocean'' of spin evolution, and the occurrence
of restricted chaos in middle of regular/periodic spin evolution. Finally, we
discuss a novel ``adiabatic resonance advection'' phenomenon, in which the
spin-orbit misalignment, trapped in a resonance, gradually evolves as
the adiabaticity parameter slowly changes. This phenomenon 
can occur for certain parameter regimes when tidal decay of the 
planetary orbit is included.
\end{abstract}


\section{Introduction}

A major surprise in exoplanetary astrophysics in recent
years is the discovery of the misalignment between the orbital axis of
the planet and the spin axis of the host star in systems containing
``hot Jupiters'', giant planets with orbital periods $\lo 5$~days
(e.g. Hebrard et al. 2008, Narita et al. 2009, Winn et al. 2009,
Triaud et al. 2010, Hebrard et al. 2010, Albrecht et al. 2012). These
planets cannot form in-situ, and must have migrated from a few AU's
distance from their host star to their current locations. Planet
migration in protoplanetary disks is usually expected to produce
aligned orbital and spin axes (however, see Bate, Lodato \& Pringle
2010; Lai, Foucart \& Lin 2011; Batygin 2012; Batygin \& Adams 2013;
Lai 2014; Spalding \& Batygin 2014). 
So the observed misalignments suggest that dynamical
interaction between planets and/or companion star may play an
important role in the planet's inward migration.

One of the dynamical channels for the migration of giant planets 
involves Lidov-Kozai oscillations (Lidov 1962; Kozai 1962) of the planet's
orbit induced by a distant companion (star or planet).
When the companion's orbit is sufficiently inclined relative to the
planetary orbit, the planet's eccentricity undergoes excursions to 
large values while the orbital axis precesses with varying inclination.
Tidal dissipation in the planet at periastron reduces the orbital energy, 
leading to inward migration and circularization of the planet's orbit 
(Wu \& Murray 2003; Fabrycky \& Tremaine 2007; Correia et al.~2011; 
Naoz et al.~2012; Petrovich 2014). A number of recent works have focused
on the extreme evolution of the planetary orbit (such as orbital flip)
when the octupole perturbing potential from the binary companion is
included (Katz, Dong \& Malhotra 2011; Naoz et al.~2011, 2013; Petrovich 2014;
see also Ford et al.~2000; Li et al.~2014; Liu, Munoz \& Lai 2014).

In a recent paper (Storch, Anderson \& Lai 2014; hereafter SAL), we
have shown that during the Lidov-Kozai cycle, gravitational
interaction between the planet and its oblate host star can lead to
complex and chaotic evolution of the stellar spin axis, depending on
the planet mass and the stellar rotation rate.  In many cases, the
variation of the stellar spin direction is much larger than the
variation of the planet's orbital axis. Moreover, in the presence of
tidal dissipation, the complex spin evolution can leave an imprint on
the final spin-orbit misalignment angle.

SAL discussed three qualitatively different regimes for the evolution of 
the spin-orbit misalignment angle $\thetaSL$. These regimes
depend on the ratio of the precession rate $\pl$
of the planetary orbital axis ($\hatL$) around the (fixed) binary axis 
$\hatLb$, and the stellar precession rate $\ps$ driven by the planet 
(see Section 2):
(i) For $|\pl|\gg |\ps|$ (``nonadiabatic'' regime),
the spin axis $\hatS$ effectively precesses around $\hatLb$, maintaining a
constant angle $\theta_{\rm sb}$ between $\hatS$ and $\hatL_b$.
(ii) For $|\ps|\gg |\pl|$ (``adiabatic'' regime),
the spin axis $\hatS$ follows $\hatL$ adiabatically as the latter
evolves, maintaining an approximately constant $\thetaSL$.
(iii) For $|\ps|\sim |\pl|$ (``trans-adiabatic'' regime), the evolution of
$\hatS$ is chaotic. However, the precise transitions between these regimes
are fuzzy.

Since both $\ps$ and $\pl$ depend on eccentricity
($\ps$ also depends on $\theta_{\rm sl}$) and thus vary strongly during 
the Lidov-Kozai cycle, a useful dimensionless
ratio that characterizes the evolution of $\hatS$ 
is the ``adiabaticity parameter'',
\be
\epsilon=\left|{\pl\over\ps}\right|_{e,\thetaSL=0},
\label{eq:epsilon1}
\ee
where the subscript implies that the quantity is evaluated
at $e=0$ and $\thetaSL=0$. 
So $\epsilon$ is constant during the Lidov-Kozai cycle.
For a planet of mass $M_p$ initially in a nearly circular 
orbit around a star of mass $M_\star$ and radius $R_\star$ 
at a semimajor axis $a$, with a binary companion of mass
$M_b$, semimajor axis $a_b$ (and eccentricity $e_b=0$), the adiabaticity parameter
is given by  
\ba
\epsilon&=&1.17
\left({k_\star\over 2k_{q}}\right)
\!\left({R_\star\over 1\,R_\odot}\right)^{\!\!-3/2}
\!\!\left({\hat\Omega_\star\over 0.1}\right)^{\!\!-1}
\!\!\left({M_b\over 10^3M_p}\right)\times \nonumber \\
&&\times
\!\left({a\over 1\,{\rm AU}}\right)^{9/2}
\!\left({a_b\over 300\,{\rm AU}}\right)^{\!-3}
\!\left|\cos\thetaLB^0\right|,
\label{epsilon}
\ea
where $\Omega_\star=(GM_\star/R_\star^3)^{1/2}\hat\Omega_\star$
is the rotation rate of the star, $k_\star/(2k_q)\sim 1$, and
$\thetaLB^0$ is the initial (at $e=0$) planetary orbital inclination
relative to the binary.
Figure 1 shows a ``bifurcation'' diagram that illustrates
the complex dynamics of the spin-orbit misalignment angle $\thetaSB$
as $\epsilon$ is varied (by changing $M_p$ while keeping other parameters
fixed). We see that in this example, wide-spread chaos occurs
for $\epsilon\go 0.14$, while the evolution of $\thetaSL$ is largely 
regular for $\epsilon\lo 0.14$. However, in the chaotic regime, 
there exist multiple periodic islands in which $\thetaSL$ evolves
regularly. Interestingly, even in the ``adiabatic'' regime, there exist
regions of ``restricted chaos'', in which $\thetaSL$ evolves chaotically but
with a restricted range.

\begin{figure*}
\includegraphics[width=\textwidth]{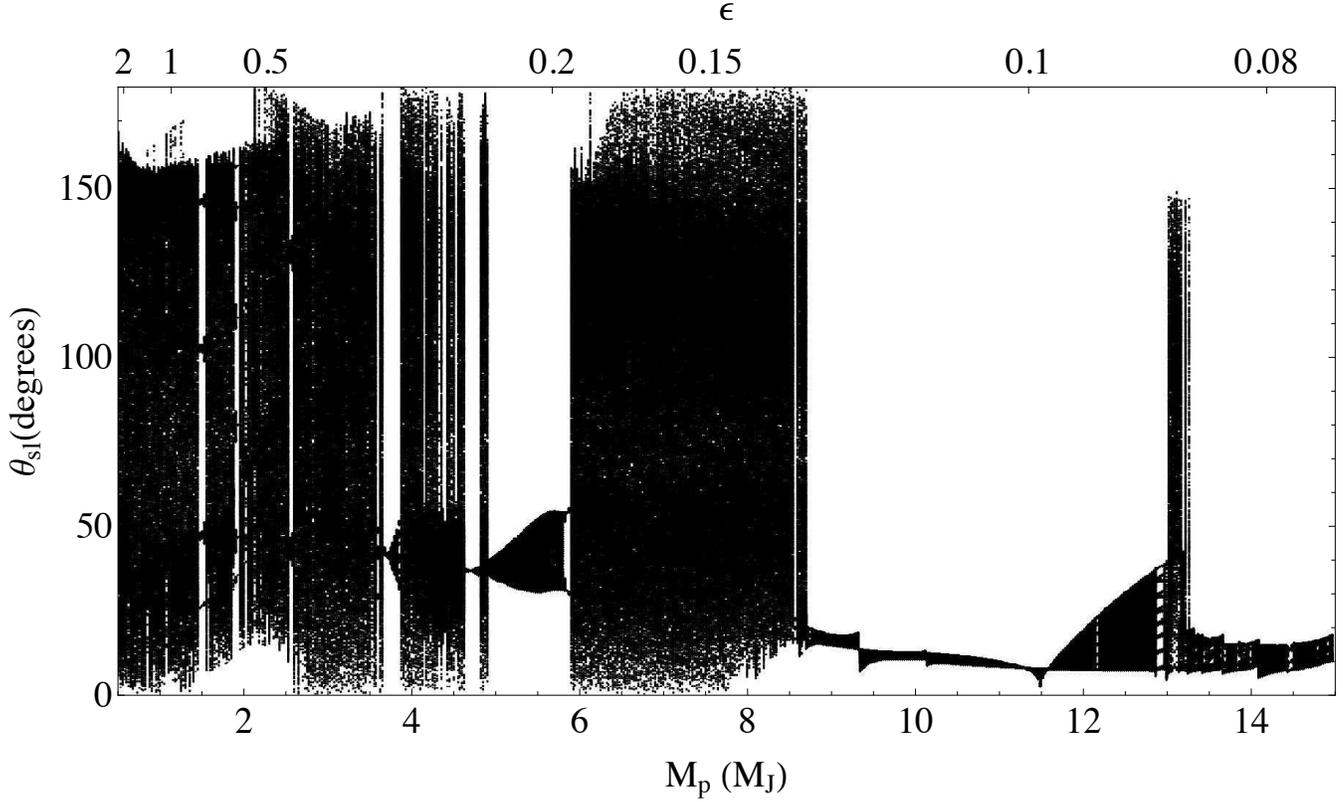}
\caption{``Bifurcation'' diagram of the spin-orbit misalignment angle
  versus planet mass and the adiabaticity parameter $\epsilon$.
 For each planet mass $M_p$, we evolve the secular orbital evolution
  equations including the effects of short-range forces (periastron
  advances due to General Relativity, the stellar quadrupole, and the planet's
  rotational bulge and tidal distortion) together with the stellar
  spin precession equation, starting with $\thetaSL=0$, for
  $\sim\!1500$ Lidov-Kozai cycles, and record $\thetaSL$ every time the
  orbital eccentricity reaches a maximum. The parameters for this plot
  are $a=1$~AU, $a_b=200$~AU, $e_0=0.01$, $\theta^0_{\rm lb} =
  85^\circ$, $\hat{\Omega}_\star=0.03$. This figure is an extended
  version of Fig.~4 of Storch, Anderson \& Lai (2014), demonstrating
  the complexity of the trans-adiabatic and even the adiabatic
  regimes of the spin dynamics.}
\label{IATfull}
\end{figure*}

Widespread chaos in dynamical systems can be understood as arising
from overlaps of resonances in the phase space (Chirikov 1979). What
are the resonances underlying the chaotic spin behaviour found in SAL
and Fig.~1? Since $\pl$ and $\ps$ are both strong functions of time,
the answer to this question is not obvious {\it a priori}, even in the
ideal case when the planetary orbit undergoes strictly periodic
Lidov-Kozai oscillations. Using Hamiltonian perturbation theory, we
show in this paper that a spin-orbit resonance occurs when the
time-averaged spin precession frequency equals an integer multiple of
the Lidov-Kozai oscillation frequency. We then demonstrate that
overlapping resonances can indeed explain the onset of chaos in the
dynamics of stellar spin, more specifically the ``adiabatic'' to
``trans-adiabatic'' transition. We also show that 
many of the intricate ``quasi-chaotic'' features found numerically in the
``adiabatic'' regime can be understood from overlapping resonances.
Finally we show that the consideration of resonances can lead to 
a novel ``adiabatic resonance advection'' phenomenon when tidal decay 
of the planetary orbit is included.

The chaotic dynamics of stellar spin studied in this paper has some
resemblance to the well-known problem of obliquity dynamics of Mars
and other terrestrial planets (Laskar \& Robutel 1993; Touma \& Wisdom
1993; see Li \& Batygin 2014).  In that problem, a spin-orbit resonance
arises when the spin precession frequency $\ps$ of Mars around its
orbital axis $\hatL$ driven by the Sun matches one of the
eigen-frequencies ($\pl$'s) for the variation of $\hatL$ due to
interactions with other planets. Only a small number of $\pl$'s are
relevant in the Solar System, and except for the $\cos\thetaSL$
factor, $\ps$ is approximately constant in time. Thus the analysis of 
overlapping resonances is relatively straightforward. For the problem
studied in this paper, by contrast, both $(\ps/\cos\thetaSL)$ and $\pl$ 
are strong functions of time, so the dynamics of the stellar spin axis
exhibits a much richer set of behaviors.

Our paper is organized as follows.  In Section 2, we review the
physical system and its ingredients.  In Section 3, we develop a
Hamiltonian formulation of the problem, and derive the resonance
condition for spin-orbit coupling.  In Section 4, we discuss the
behaviour of the system under the influence of a single resonance. In
Section 5, we demonstrate the onset of chaos in the presence of two or
more overlapping resonances, and derive the overlap criterion.  In
Section 6, we consider the full Lidov-Kozai driven spin precession
problem, and demonstrate that resonance overlaps can explain the onset
of chaos, as well as other ``quasi-chaotic'' features in the spin
evolution. In Section 7, we consider the effect of a slowly evolving
adiabaticity parameter, as a simplified model of tidal dissipation,
and present a proof of concept for understanding the novel ``adiabatic resonance
advection'' phenomenon. We summarize our key findings in Section 8.

\section{Review of the Physical System and Ingredients}

\subsection{Lidov-Kozai (LK) Oscillations}

We consider a planet of mass $M_p$ in orbit around a host star of mass
$M_\star$ (with $M_\star\gg M_p$), and a distant companion of mass
$M_b$.  The host star and companion are in a static orbit with
semi-major axis $a_b$, eccentricity $e_b$,
and angular momentum axis $\hatLb$,
which defines the invariant plane of the system.
The planet's orbit has semi-major axis $a$, eccentricity $e$, 
angular momentum axis $\hatL$ and inclination $\thetaLB$ (the angle 
between $\hatL$ and $\hatLb$). 
In the Lidov-Kozai (LK) mechanism, the quadrupole potential of the companion
causes the orbit of the planet to undergo oscillations of both $e$ and
$\thetaLB$, as well as nodal precession ($\dot{\Omega}$) and
pericenter advance ($\dot{\omega}$), while conserving 
${\bf L}\cdot\hatLb$. The equations governing these oscillations 
are given by
\ba
{de\over dt}&=& t_k^{-1} {15\over 8}e\sqrt{1-e^2}\sin2\omega\sin^2\thetaLB, \label{ek} \\
{d\Omega\over dt}&=& t_k^{-1} {3\over 4}{\cos\thetaLB\left(5e^2\cos^2\omega-4e^2-1\right)\over\sqrt{1-e^2}}, \label{Omk} \\
{d\thetaLB\over dt}&=&- t_k^{-1} {15\over 16}{e^2\sin2\omega\sin2\thetaLB\over\sqrt{1-e^2}}, \label{ik} \\
{d\omega\over dt}&=& t_k^{-1}{3\left[2(1-e^2)+5\sin^2\omega(e^2-\sin^2\thetaLB)\right]\over 4\sqrt{1-e^2}}, \label{omk}
\ea
where $t_k^{-1}$ is the characteristic frequency of oscillation, given by
\be
t_k^{-1}={n\over (1-e_b^2)^{3/2}}
\left(\frac{M_b}{M_\star}\right)\left(\frac{a}{a_b}\right)^3,
\ee 
where $n=\sqrt{GM_\star/a^3}$ is the planet's mean motion.
In this paper, we neglect all effects associated with short-range forces
(General Relativity, tidal interaction, etc) and the octupole potential
from the binary.

Equations (\ref{ek})-(\ref{omk}) admit two types of analytical
solutions, distinguished by whether the argument of pericenter
$\omega$ circulates or librates. In the present work we will consider
only the circulating case by taking $\omega=0$ at $t=0$. The conservation
of the projected angular momentum ${\bf L}\cdot\hatLb$ gives
\be
x\cos^2\thetaLB=x_0\cos^2\thetaLB^0\equiv h,
\label{eq:kozai}\ee
where
\be
x\equiv 1-e^2,
\ee
and energy conservation gives
\be
e^2(5\sin^2\thetaLB\sin^2\omega-2)=-2e_0^2.
\label{eq:energy}\ee
For the initial eccentricity $e_0\approx 0$, the above equations imply that 
the maximum eccentricity occurs at $\omega=\pi/2,\,3\pi/2$, where
$\sin^2\thetaLB=2/5$ and 
\be
e_{\rm max}\simeq \left(1-{5\over 3}\cos^2\thetaLB^0\right)^{1/2}.
\ee

Combining eqs.~(\ref{eq:kozai})-(\ref{eq:energy}) with eq.~(\ref{ek}),
the time evolution of eccentricity can be solved explicitly
(Kinoshita \& Nakai 1999):
\be
x=x_0 + (x_1-x_0){\rm cn}^2(\theta,k^2), \label{x}
\ee
where
\ba
\theta&=&\frac{K}{\pi}\left(n_e t +\pi\right), \\
n_e &=& t_k^{-1}\frac{6\pi\sqrt{6}}{8K}\sqrt{x_2-x_1}, \\
k^2 &=& \frac{x_0-x_1}{x_2-x_1}.
\ea
In the above expressions ${\rm cn}(\theta,k^2)$ is the Jacobi elliptic
${\rm cn}$ function with modulus $k^2$, $n_e$ is the ``mean motion'' for the 
eccentricity variation (i.e. $2\pi/n_e$ is the period of the eccentricity oscillations), 
$K$ is the complete elliptic
integral of the first kind with modulus $k^2$, $x_0$ is 
the value of $x$ at $t=0$, and $x_1$ and $x_2$ ($x_1 <
x_2$) are solutions to the quadratic equation
\be
x_{1,2}^2-\frac{1}{3}\left(5+5h-2x_0\right)x_{1,2}+\frac{5}{3}h=0,
\ee
obtained from eqs.~(\ref{eq:kozai})-(\ref{eq:energy}) with $\sin^2\omega=1$.
The other orbital elements can be
expressed as a function of $x$. Note that the period of $\omega$ circulation 
($\omega$ goes from 0 to $2\pi$) is $4\pi/n_e$.

For the remainder of this work, we use a single $x(t)$ solution in our
analysis, 
corresponding to $e_0=0.01$ (so $x_0 =1-(0.01)^2$) and $\theta^0_\lb=85^\circ$.

\subsection{Stellar spin precession}

Because of the rotation-induced oblateness, the star is 
torqued by the planet, causing its spin
axis $\hatS$ to precess around the planet's orbital axis $\hatL$
according to the equation
\be
\frac{d\hatS}{dt} = \ps \hatL\times\hatS,
\label{dsdt}
\ee
where the precession frequency $\ps$ is given by 
\be
\ps= -\frac{3GM_p(I_3-I_1)}{2a^3(1-e^2)^{3/2}}\frac{\cos\thetaSL}{S}.
\label{omps}
\ee
Here $I_3$ and $I_1$ are the principal moments of inertia of the star, $S$
is the magnitude of the spin angular momentum, and $\thetaSL$ is the
angle between $\hatS$ and $\hatL$. 
Our goal is to characterize how
$\thetaSL$ changes as a function of time as the planet's orbit
undergoes LK oscillations. 
Since $e$ changes during the LK cycle, we write
the spin precession frequency as
\be
\ps(t)\equiv -\alpha(t)\cos\thetaSL = -\frac{\alpha_0}{x^{3/2}}\cos\thetaSL,
\ee
where
\ba
\alpha_0
&=&{3GM_p(I_3-I_1)\over 2a^3 I_3\Omega_\star}\nonumber\\
&=&1.19\times10^{-8}\left({2\pi\over1\mathrm{yr}}\right)\left({2k_{q}\over k_\star}\right)\left({10^3M_p\over M_\star}\right)\left({\hat{\Omega}_\star\over0.05}\right)\times \nonumber \\
&&\times \left({a\over1\mathrm{AU}}\right)^{-3}\left({M_\star\over M_\odot}\right)^{1/2}\left({R_\star\over R_\odot}\right)^{3/2}.
\label{alpha0}
\ea
Here we have used $(I_3-I_1)\equiv k_q
M_\star R_\star^2\hat{\Omega}_\star^2$, 
with $\hat\Omega_\star=\Omega_\star/(GM_\star/R_\star^3)^{1/2}$ the dimensionless
stellar rotation rate, and $S=I_3\Omega_s\equiv k_\star M_\star
R_\star^2\Omega_\star$. 
For a solar-type star, $k_q\approx 0.05$, and
$k_\star\approx 0.1$ (Claret 1992).

During the LK cycle, the planet's orbital axis $\hatL$ 
changes in two distinct ways: nodal precession around $\hatL_b$
at the rate $\pl(t)=\dot{\Omega}$, and nutation at the rate 
$\dot{\theta}_{\rm lb}(t)$. 
Each of these acts as a driving force for the stellar spin.
The variation of $\thetaLB(t)$ plays an important role as well since
it affects $\hatL(t)$ directly [see Eq.~(\ref{eq:hal}) below].
Note that the back-reaction torque from the stellar quadrupole on the orbit
also acts to make $\hatL$ precess around $\hatS$; we neglect this back-reaction
throughout this paper in order to focus on the spin dynamics with ``pure''
orbital LK cycles.
Based on the analytical LK solution given in the previous sub-section, 
we find $\pl$ is given by 
\be
\pl = \dot\Omega=\Omega_{\rm pl,0} \left[1-\frac{2(x_0-h)}{x-h}\right],
\label{pl}
\ee 
with $x$ is given by Eq.~(\ref{x}) and
\be
\Omega_{\rm pl,0}=\frac{3}{4t_k}\sqrt{h} \simeq \frac{3}{4 t_k}\left|\cos\theta^0_{\rm lb}\right|,
\ee
where the second equality assumes $e_0\simeq 0$.
The angle $\thetaLB$ and its derivative are given by $\cos\thetaLB=\sqrt{h/x}$ and 
$\dot{\theta}_{\rm lb} = \dot{x}\cos\thetaLB/(2x\sin\thetaLB)$. Note
that $\pl(t) < 0$. The quantity $\Omega_{\rm pl,0}$ specifies the value of $\left|\pl\right|$ at $e=e_0\simeq 0$, and is explicitly given by 
\ba
\Omega_{\rm pl,0}&\simeq&\frac{3}{4}\left({2\pi\over 10^6\mathrm{yr}}\right)\left({M_b\over M_\star}\right)\left({M_\star\over M_\odot}\right)^{1/2}\left({a\over 1\mathrm{AU}}\right)^{3/2}\times \nonumber \\
&&\times\left({a_b\over 100\mathrm{AU}}\right)^{-3}{\left|\cos\theta^0_{\rm lb}\right|\over(1-e_b^2)^{3/2}},
\label{pl0}
\ea 
for $x_0 = 1-e_0^2\simeq 1$. Taking the ratio of this and
Eq. (\ref{alpha0}) yields the adiabaticity parameter
\be
\epsilon={\Omega_{\rm pl,0}\over\alpha_0}, 
\label{eq:epsilon}\ee
as given in Section 1 [see Eq.~(\ref{epsilon})].

\section{Hamiltonian Formulation of Spin Dynamics and Resonances}

\subsection{The Spin Hamiltonian}
 
In the inertial frame, the Hamiltonian governing the dynamics of stellar
spin ${\bf S}=S\hatS$ is
\be
H={S^2\over 2I_3}+{GM_p(I_3-I_1)\over 4a^3 (1-e^2)^{3/2}}\left[1-3(\hatS\cdot\hatL)^2
\right].
\label{eq:Hinert}\ee
The first term is the (constant) rotational kinetic energy and will be dropped
henceforth, and the second term is the orbital-averaged interaction energy between 
the planet and stellar quadrupole. 
Since the evolution of the orbital eccentricity $e(t)$ is fixed, we only need 
to consider the last term in Eq.~(\ref{eq:Hinert}):
\be
H_0\equiv -{1\over 2}\alpha(t)S\left(\hatS\cdot\hatL\right)^2.
\ee
Noting that $\hatS\cdot\hatL_b$ and $\phi_s$ (the precessional phase of 
$\hatS$ around $\hatL_b$) are conjugate variables, 
we can check that the Hamiltonian equations for $H_0$ lead to
Eq.~(\ref{dsdt}).

Since we are interested in the variation of $\thetaSL$, it is
convenient to work in the rotating frame in which $\hatL$ is a
constant.  In this frame, the Hamiltonian takes the form
(cf. Kinoshita 1993)
\be
H_{\rm rot}=H-{\bf R}\cdot {\bf S},
\ee
where the rotation ``matrix'' is 
\be
{\bf R}=\pl\hatLb+ {\dot\theta}_{\rm lb} \left({\hatLb\times\hatL\over\sin\thetaLB}
\right).
\ee
To write down the explicit expression for $H_{\rm rot}$, we set up
a Cartesian coordinate system with the $z$-axis along $\hatL$, and 
the $x$-axis pointing to the ascending node of the planet's orbit in the
invariant plane (the plane perpendicular to $\hatLb$). The spin axis 
is characterized by $\thetaSL$ and the precessional phase $\phi$ (the longitude
of the node of the star's rotational equator in the $xy$-plane), such that 
\be
\hatS=\sin\thetaSL \left(\sin\phi \,{\hat x}-\cos\phi \,{\hat y}\right)+\cos\thetaSL
\,{\hat z}.
\ee
Setting $S=1$ and suppressing the subscript ``rot'', we have
\ba
&&H= -{1\over 2}\alpha(t)\left(\cos\thetaSL\right)^2 
-\dot{\theta}_{\rm lb}(t)\sin\thetaSL\sin\phi 
\nonumber\\ 
&&\quad -\pl(t)\Bigl[\cos\thetaLB(t) \cos\thetaSL - \sin\thetaLB(t) \sin\thetaSL\cos\phi
\Bigr],
\label{eq:hal}\ea
Note that $\phi$ and $\cos\thetaSL$ are the conjugate pair of
variables we wish to solve for. Since in this work we focus on the
behavior of the system close to the adiabatic regime, in general the
first term in the Hamiltonian dominates, while the others can be
treated as perturbations. In the limit of no perturbation, the zeroth
order Hamiltonian $H_0\equiv -{1\over 2}\alpha(t)\cos^2\thetaSL$ indeed
conserves $\cos\thetaSL$, as it should based on the arguments given in
Section 1.

\begin{figure}
\scalebox{0.39}{\includegraphics{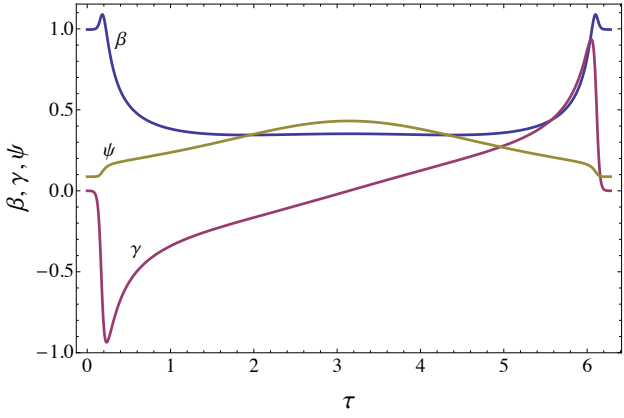}}
\caption{Plots of the ``shape'' functions $\beta(\tau)$ (blue),
  $\gamma(\tau)$ (red), and $\psi(\tau)$ (brown), for $x_0=1-0.01^2$ and
  $\cos\theta^0_{\rm lb}=85^\circ$.}
\label{betagammapsi}
\end{figure}

\subsection{The Renormalized Hamiltonian}

The equations of motion for $\phi$ and $\cos\thetaSL$ can be derived
from the Hamiltonian (\ref{eq:hal}), and are given by
\ba
&&\frac{d\phi}{dt} =
-\alpha(t)\cos\thetaSL+\dot{\theta}_{\rm lb}(t)\frac{\cos\thetaSL}{\sin\thetaSL}\sin\phi \nonumber\\
&&\qquad\quad
-\pl(t)\left[\cos\thetaLB(t) + \sin\thetaLB(t) \frac{\cos\thetaSL}{\sin\thetaSL}\cos\phi\right],\\
&&\frac{d\cos\thetaSL}{dt} = \pl(t)\sin\thetaLB(t)\sin\thetaSL\sin\phi 
\nonumber\\
&&\qquad\quad + \dot{\theta}_{\rm lb}(t)\sin\thetaSL\cos\phi.
\ea
These equations can be simplified by introducing a rescaled time
variable $\tau$ such that $d\tau \propto \alpha(t) dt$, i.e.,
\be
\tau(t) = \frac{n_e}{\bar{\alpha}} \int_0^t \alpha(t') dt', 
\ee
where
\be
\bar{\alpha}\equiv\frac{n_e}{2\pi}\int_0^{2\pi/n_e}\!\alpha(t)dt.
\ee
Here the factor of $n_e/\bar{\alpha}$ is used to ensure that all of
the time-dependent forcing functions introduced in Section 2 have
a period of $2\pi$ in $\tau$-space, for convenience. The equations of
motion in $\tau$ space are then given by
\ba
&&\frac{d\phi}{d\tau} =\frac{\bar{\alpha}}{n_e}\biggl\{-\cos\thetaSL
+\frac{\dot{\theta}_{\rm lb}(\tau)}{\alpha(\tau)}\frac{\cos\thetaSL}{\sin\thetaSL}
\sin\phi \nonumber\\
&& \quad ~~ -\frac{\pl(\tau)}{\alpha(\tau)}\Bigl[\cos\thetaLB(\tau) 
+ \sin\thetaLB(\tau) \frac{\cos\thetaSL}{\sin\thetaSL}\cos\phi\Bigr]\!\biggr\},\\
&&\frac{d\cos\thetaSL}{d\tau} = \frac{\bar{\alpha}}{n_e}\biggl\{\frac{\pl(\tau)}
{\alpha(\tau)}\sin\thetaLB(\tau)\sin\thetaSL\sin\phi \nonumber\\
&&\quad ~~ + \frac{\dot{\theta}_{\rm lb}(\tau)}{\alpha(\tau)}\sin\thetaSL\cos\phi
\biggr\}.
\ea
The corresponding Hamiltonian is
\ba
&&H'(p,\phi,\tau)=\frac{\bar{\alpha}}{n_e}\biggl\{
-\frac{1}{2}p^2 +\epsilon \psi(\tau)\,p \nonumber\\
&&\qquad ~~-\epsilon \sqrt{1-p^2}\,\Bigl[\beta(\tau)\cos\phi
+\gamma(\tau)\sin\phi\Bigr]\biggr\}, 
\label{kozaiham}\ea
where we have defined $p\equiv \cos\thetaSL$, and
\ba
&&\epsilon\beta(\tau)=-\frac{\pl(\tau)}{\alpha(\tau)}\sin\thetaLB(\tau), \label{beta} \\
&&\epsilon\gamma(\tau)=\frac{\dot{\theta}_{\rm lb}(\tau)}{\alpha(\tau)}, \label{gamma}\\
&&\epsilon\psi(\tau)=-\frac{\pl(\tau)}{\alpha(\tau)}\cos\thetaLB(\tau) \label{psi}.
\ea
Since $\epsilon={\Omega_{\rm pl,0}/\alpha_0}$ [see Eq.~(\ref{eq:epsilon})],
the functions $\beta(\tau)$, $\gamma(\tau)$ and $\psi(\tau)$ depend
only on the ``shape'' of the orbit, i.e., on $e(\tau)$ (with $\tau$ varying from
0 to $2\pi$). For a given $\thetaLB^0$ (and $e_0\simeq 0$), these functions are fixed
and do not depend on any other parameters.
Figure \ref{betagammapsi} depicts these functions for $\thetaLB^0=85^\circ$.

\subsection{Fourier Decomposition and Resonances}

\begin{figure}
\scalebox{0.56}{\includegraphics{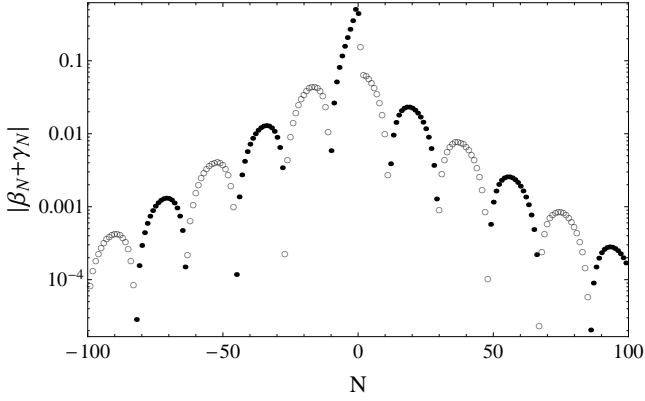}}
\caption{Sum of the Fourier coefficients as a function of the
  resonance number $N$. Filled circles are for positive
  $\beta_N+\gamma_N$, and open circles are for negative
  $\beta_N+\gamma_N$. Note that $\beta_{-N}=\beta_N$, while
  $\gamma_{-N}=-\gamma_N$, which accounts for the lack of symmetry
  across $N=0$.}
\label{coeffs}
\end{figure}

We now expand $\beta(\tau), \gamma(\tau)$, and $\psi(\tau)$ in Fourier
series. Since $\beta$ and $\psi$ are symmetric with respect to $\tau=\pi$,
while $\gamma$ is anti-symmetric (see Fig.~\ref{betagammapsi}), we have
\ba
&&\beta(\tau)=\sum_{M=0}^{\infty} \beta_M \cos M\tau, \\
&&\gamma(\tau)=\sum_{M=1}^{\infty} \gamma_M \sin M\tau, \label{gamma}\\
&&\psi(\tau) =\sum_{M=0}^{\infty} \psi_M \cos M\tau.
\ea
Obviously, $\beta_M$, $\gamma_M$ and $\psi_M$ depend only on the
``shape'' of the orbit $e(\tau)$.
The Hamiltonian (\ref{kozaiham}) becomes
\ba
&&H'=\frac{\bar{\alpha}}{n_e} \biggl\{-\frac{1}{2}p^2 
+ \epsilon\,\psi_0\, p + \epsilon\, p \sum_{M=1}^{\infty} \psi_M \cos Mt \nonumber \\
&&\qquad ~-\frac{\epsilon}{2}\sqrt{1-p^2}\sum_{M=0}^{\infty}\Bigl[
(\beta_M+\gamma_M)\cos(\phi- M\tau) \nonumber \\
&&\qquad\quad\quad\quad\quad+(\beta_M-\gamma_M)\cos(\phi+M\tau)\Bigr]\biggr\}. 
\label{hexp}
\ea
Note that $\gamma_0$ is not defined in Eq.~(\ref{gamma}). For
convenience of notation, we will set $\gamma_0=\beta_0$ [see
discussion following Eq.~(\ref{oneharmonic})].

A resonance occurs when the argument of the cosine function, $(\phi\pm
M\tau)$, in the Hamiltonian (\ref{hexp}) is slowly varying, i.e., when 
${d\phi}/{d\tau} = N$, where $N$ is a positive or negative integer. 
In the perturbative regime ($\epsilon \ll 1$) of interest in this paper, 
the Hamiltonian is dominated by $H_0 = (\bar{\alpha}/n_e)(-p^2/2)$, and we have
${d\phi}/{d\tau}\simeq -\bar{\alpha}p/n_e$. So the resonance condition
becomes
\be
{\bar\Omega}_{\rm ps}
=-\bar{\alpha}\cos\thetaSL = N n_e,\qquad {\rm with}~~N = 0,\pm 1,\pm 2,\pm 3,\cdots
\label{rescond}
\ee
i.e. the averaged stellar precession frequency ${\bar\Omega}_{\rm ps}$ 
equals an integer multiple of the mean eccentricity oscillation frequency in the
LK cycle. Note that, since $\cos\thetaSL$ spans the range
$\{-1,1\}$, this means that for any given value of $\bar{\alpha}$ and
of $n_e$ there exist multiple resonances. We may then define the
zeroth-order resonant momentum corresponding to each resonance as
\be
p_N=\left(\cos\theta_{\rm sl}\right)_N = -\frac{N n_e}{\bar{\alpha}}.
\label{pr}
\ee
Since $|p_N|$ cannot exceed $1$, we also see that there exists a 
``maximum resonance order'',
\be
N_{\rm max} = \left\lfloor\frac{\bar{\alpha}}{n_e}\right\rfloor =\left\lfloor\frac{1}{\epsilon}\mathcal{N}\left(\cos\theta^0_{\rm lb};e_0\right)\right\rfloor,
\label{nmax}
\ee
such that $N=N_{\rm max}$ is the maximum allowed positive resonance,
and $N=-N_{\rm max}$ is the maximum allowed negative resonance. 
Note that the resonant momentum $p_N$ can be written as 
\be
p_N\simeq -{N\over N_{\rm max}}.
\label{eq:pn}\ee
Thus, the stellar spin evolution is perturbed by a set of $(2N_{\rm max} + 1)$ 
resonances. The function $\mathcal{N}$ depends mainly on $\cos\theta^0_{\rm lb}$,
and weakly on $e_0$ (assuming $e_0 \ll 1$). For 
$\theta^0_{\rm lb}=85^\circ$ (adopted for our numerical examples in this paper), 
we find $\mathcal{N}=0.98$. 

The $N=-N_{\rm max}$ resonance is of particular interest, as it is the
closest resonance to $p_N = 1$, the aligned configuration. Thus,
if a star-planet system is born with the stellar spin axis and the
planet orbital axis aligned,
this resonance is the one that most directly influences the stellar spin evolution
This will be discussed in detail in Section 6.

We may now ask what happens if the resonance condition is satisfied:
how are the dynamics of stellar spin precession affected by one - or
more - resonances? To make the solution tractable analytically, we
must make some simplifying assumptions. We assume $\epsilon$ is small,
i.e. the system is in or close to the adiabatic regime. As a
corollary, we assume that individual resonances do not affect each
other significantly, i.e., that we may analyze the resonances one at a
time rather than consider the coupling between them.

\begin{figure}
\scalebox{0.55}{\includegraphics{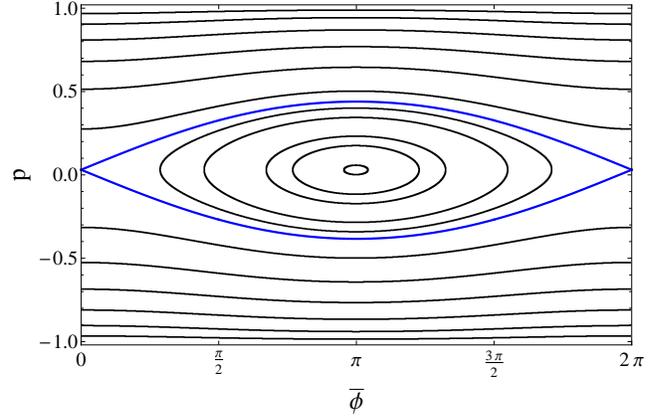}}
\caption{Sample constant-energy curves for the $N=0$ single-resonance
  Hamiltonian [given by Eq.~(\ref{oneharmonic})], constructed by
  starting with a variety of initial conditions (corresponding to
  unique values of E) and evolving the equations of motion derived
  from the Hamiltonian. The blue line shows the analytical prediction
  for the separatrix. The adiabaticity parameter is $\epsilon=0.1$,
  i.e. $\alpha_0 = 10\Omega_{\rm pl,0}$.}
\label{N0}
\end{figure}

\section{Dynamics of a Single Resonance}

To examine the dynamics of a particular single resonance (labeled by $N$,
which can be either positive or negative), it is useful to transform the 
Hamiltonian into the frame of reference in which that resonance is stationary. 
To this end, we perform a canonical transformation to the new coordinates
$(\bar{\phi},\bar{p})$ such that $\bar{\phi} = \phi - N\tau$. Using the 
generating function $F_2 = (\phi-N\tau)\bar{p}$, we then find 
\ba
&& \bar{p}=p, \quad \bar{\phi} = \phi - N\tau,\\
&& \bar{H'} = H'\left[\phi(\bar{\phi}),p(\bar{p});\tau\right] - N\bar{p}.
\ea
Thus the transformed Hamiltonian is 
\ba
&&\bar{H'}=\frac{\bar{\alpha}}{n_e}\biggl\{-\frac{1}{2} p^2 
-\frac{n_e}{\bar{\alpha}}Np  
+\epsilon\, p\sum_{M=0}^{\infty} \psi_M \cos M\tau  \nonumber\\
&&\qquad - \frac{\epsilon}{2}\sqrt{1-p^2}\sum_{M=0}^{\infty}\Bigl[
(\beta_M+\gamma_M)\cos\left[\bar{\phi}- (M-N)\tau\right] \nonumber \\
&& \quad\quad\quad\quad\quad +(\beta_M-\gamma_M)\cos
\left[\bar{\phi}+(M+N)\tau\right]\Bigr]\biggr\}, 
\ea
where we have dropped the bar over $p$, since $\bar{p}=p$. We now take the average
\be
\tilde{H}_N = \frac{1}{2\pi}\int_0^{2\pi}\bar{H'}d\tau.
\ee
Note that all the terms in the sums are rapidly varying and are averaged out
except the $M=0$ term in the first sum, the $M=N$ term (when
$N>0$) in the second sum, and/or the $M=-N$ term (when $N<0$) in the third sum.
We then have
\ba
&& \bar{H}_N = \frac{\bar{\alpha}}{n_e}\biggl[-\frac{1}{2}p^2 
-\frac{n_e}{\bar{\alpha}}Np \nonumber\\
&&\qquad +\epsilon\,\psi_0 \,p  - \frac{\epsilon}{2}\sqrt{1-p^2}\,
(\beta_N+\gamma_N)\cos\bar{\phi}\biggr],
\label{oneharmonic}
\ea
where we have used $\beta_{-N}=\beta_{N}$ and $\gamma_{-N}=-\gamma_N$. 
In order to ensure that this expression is valid for all $N$'s (including
$N=0$), we set $\gamma_0 = \beta_0$. 

The Hamiltonian (\ref{oneharmonic}) shows that the sum of Fourier
coefficients $(\beta_N+\gamma_N)$ plays a
key role in determining the property of the $N$-resonance.
Figure \ref{coeffs} plots $\left(\beta_N + \gamma_N\right)$ versus $N$, showing 
that it oscillates from positive to negative in a ringdown fashion.
This oscillatory behaviour arises from individual 
ringdowns in $\beta_N$ and $\gamma_N$, as well as from 
interference between the $\beta_N$ and $\gamma_N$ terms.

\begin{figure}
\scalebox{0.55}{\includegraphics{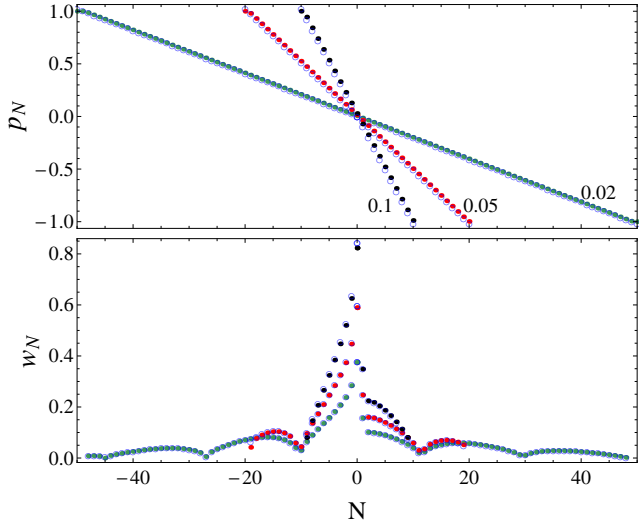}}
\caption{Comparison of the exact resonance locations (upper panel) and
  widths (bottom panel) obtained by solving Eq.~(\ref{oneharmonic})
  (filled circles) with simple analytical estimates (open
  circles). Three different values of $\epsilon$ are considered:
  $\epsilon=0.1$ (blue points), $\epsilon=0.05$ (red points), and
  $\epsilon=0.02$ (green points). The agreement between the exact
  calculation and simple estimates is quite good, and gets better with
  smaller $\epsilon$.}
\label{widths}
\end{figure}

\begin{figure}
\scalebox{0.6}{\includegraphics{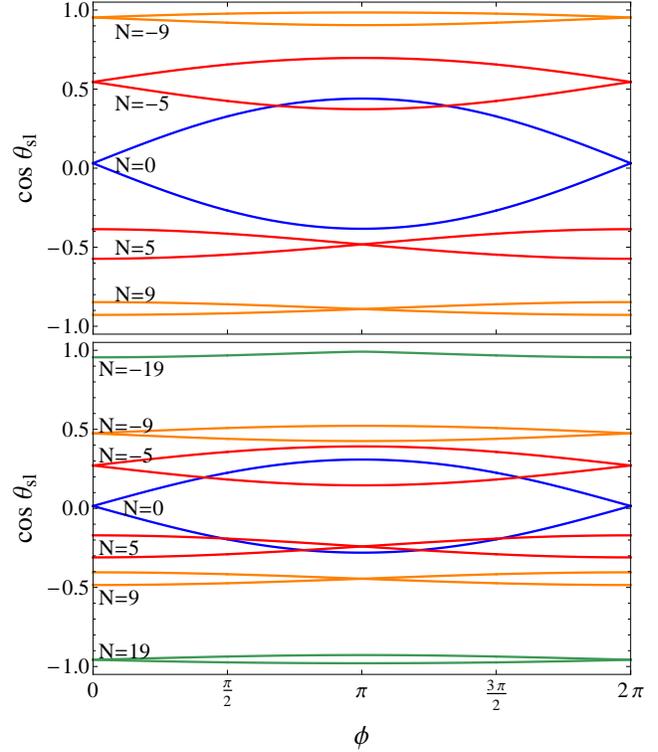}}
\caption{Sample separatrices for resonances of different $N$'s, for
  $\epsilon=0.1$ (top) and $\epsilon=0.05$ (bottom). The top panel has
  $N_{\rm max} = 9$, and the bottom panel has $N_{\rm max}=19$.}
\label{Nmult}
\end{figure}

Since the Hamiltonian (\ref{oneharmonic}) 
is not explicitly dependent on time, energy conservation holds, i.e.
\be
\bar{H}_N(\bar{\phi},p)=\bar{H}_N(\bar{\phi}_0,p_0)\equiv E,
\ee
for a trajectory that starts at $(\bar{\phi}_0,p_0)$. This 
equation is quartic which can be solved for $p(\bar{\phi};E)$. 
Figure \ref{N0} shows the constant-energy curves in the phase
space for $N=0$, illustrating the major features of the solution. The
trajectories come in two distinct flavors: those that {\it circulate},
i.e. cover the entire range of $\bar{\phi}$ and do not cross 
$p=p_N$ [see Eq.~(\ref{pr})],
and those that {\it librate}, i.e. are confined to some limited
range of $\bar{\phi}$. 
The center of the librating island is the true
location of the resonance, which is a stable fixed point of the
equations of motion. Separating the librating and circulating regions
of the phase space is a special curve known as the separatrix, which
connects two saddle fixed points. The width of the separatrix (in the $p$
axis) defines the width of the resonance.

To derive a simple expression for the resonance width, we may
simplify the Hamiltonian (\ref{oneharmonic}) further by expanding
it around $p=p_N$, where $p_N$ is the zeroth-order resonant momentum
given by Eq. (\ref{pr}). We take $p=p_N + \delta p$, assume the terms
proportional to $\epsilon$ are already small, and expand 
Eq.~(\ref{oneharmonic}) to second order in $\delta p$:
\be
\bar{H}_N \simeq \frac{\bar{\alpha}}{n_e}\biggl[
-\frac{1}{2}\delta p^2 - \frac{\epsilon}{2}\sqrt{1-p_N^2}\,(\beta_N+\gamma_N)
\cos\bar{\phi}\biggr],
\label{onehapprox}
\ee  
where constant terms (which do not depend on $\delta p$) have been dropped.
Equation (\ref{onehapprox}) is the Hamiltonian of a simple Harmonic oscillator.
The resonance width is given by
\be
w_N\simeq 2\left[2\,\epsilon\, |\beta_N+\gamma_N|\sqrt{1-p_N^2}\right]^{1/2}.
\label{appwidth}
\ee

Figure \ref{widths} shows a comparison of the exact locations of the
resonances (the fixed points of Eq.~\ref{oneharmonic})
\footnote{ Note that, in general, Eq.~\ref{oneharmonic} admits several
  fixed points. Besides the resonance fixed point $p=p_N$, other fixed
  points exist at values of $p$ very close to $\pm 1$. However, these
  fixed points do not globally affect the system; their separatrices
  are very localized. The limited influence of one such fixed point
  can be seen in Fig.~\ref{fullvnmax} (left) for $p\approx 1$.}  with
the unperturbed value $p_N$ [see Eq.~(\ref{pr})], as well as a
comparison of the exact widths of the resonances with
Eq.~(\ref{appwidth}).  We see that the approximate Hamiltonian
(\ref{onehapprox}) reproduces the resonance properties of the full
Hamiltonian (\ref{oneharmonic}) accurately.  Note that the resonance
width depends on the sum of fourier coefficients $|\beta_N+\gamma_N|$,
and since $\beta$ is symmetric while $\gamma$ is antisymmetric with
respect to $N$, the positive and negative resonances do not have the
same widths. Furthermore, since $\left(\beta_N+\gamma_N\right)$ goes
through zero several times in the interval $N\in\{-100,100\}$, the
resonance width is non-monotonic as a function of $N$.

\begin{figure}
\scalebox{0.6}{\includegraphics{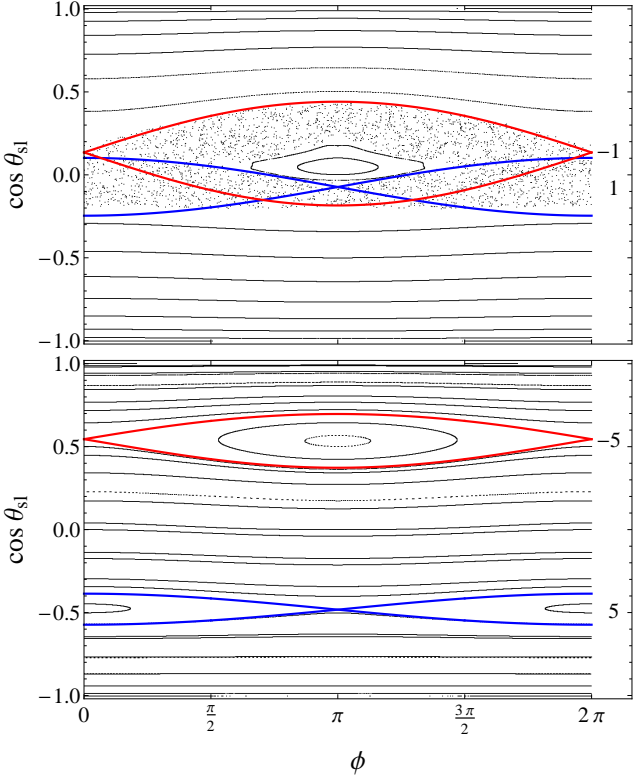}}
\caption{Surfaces of section for two different pairs of
  resonances. {\it Top panel}~: $N=1,\,M=-1$. {\it Bottom panel}~: $N=5,\,M=-5$.  The
  adiabaticity parameter is $\epsilon=0.1$. The red and blue curves in
  each panel show the analytically computed separatrices for each of
  the resonances, using the method of Section 4 (i.e. each resonance
  is analyzed separately).}
\label{cnoc}
\end{figure}

Figure \ref{Nmult} shows several separatrices for resonances of different
orders (i.e. different $N$s) obtained by solving the full Hamiltonian 
(\ref{oneharmonic}), for two different values of $\epsilon$. 
(Note we vary $\epsilon$ by varying $\alpha_0$ while
keeping $\Omega_{\rm pl,0}$ fixed;
this means that the ``shape'' functions are unchanged.) 
Figure \ref{Nmult} illustrates several
different features of the separatrices. 
First, decreasing $\epsilon$ tends to decrease the displacement of
individual resonances from $p=0$. Each resonance is centered at
$p\simeq p_N\simeq -N/N_{\rm max}$. 
Since the maximum order of resonance, $N_{\rm max}$
(recall that no resonance is possible for $|N|>N_{\rm max}$; see Section 3.3),
is inversely proportional to $\epsilon$ (Eq.~\ref{nmax}), we have 
$|p_N|\propto \epsilon$.
Second, the general trend is that at smaller $\epsilon$ all the resonances 
are narrower, though this is not precisely true 
because $p_N$ also depends on $\epsilon$ [see Eq.~(\ref{appwidth})].
Finally, the position of the resonance in the $\phi$ coordinate depends on the sign of
$(\beta_N+\gamma_N)$: if $(\beta_N+\gamma_N)<0$, the resonance is
located at $\phi=\pi$, and if $(\beta_N+\gamma_N)>0$ -- at $\phi=0$. 
Since $\gamma_{-N} = -\gamma_N$, this usually implies that there are
significant differences between resonances with $N>0$ and those with
$N<0$.

To summarize, given a particular value of the adiabaticity parameter
$\epsilon$, the stellar spin is perturbed by a set of
resonances $d\phi/d\tau = N$ with $N\in\{-N_{\rm max},N_{\rm max}\}$,
where $N_{\rm max}$ is given by Eq.~(\ref{nmax}). Each resonance
governs the stellar spin evolution in the vicinity of $\cos\thetaSL =
p_N$, with $p_N$ approximately given by Eq.~(\ref{pr}), and the width of 
the governed region approximately given by Eq.~(\ref{appwidth}). As
$\epsilon$ decreases (the system becomes more adiabatic), 
$N_{\rm max}$ increases, $|p_N|\simeq |N|/N_{\rm max}$ (for a given $N$)
decreases (the resonance locations move closer to $p=0$), and the
width of the resonance generally decreases. 
For a given $\epsilon$, the width of the resonance is a non-monotonic
function of $N$ because of its dependence on $(\beta_N+\gamma_N)$.

\section{Onset of Chaos: Two or More Resonances}

We now consider a Hamiltonian of the form 
\ba
&& H =\frac{\bar{\alpha}}{n_e}\biggl\{
-\frac{1}{2} p^2 +\epsilon\,\psi_0\, p \nonumber \\
&&\qquad - \frac{\epsilon}{2}\sqrt{1-p^2}\Bigl[(\beta_N+\gamma_N)\cos(\phi-N\tau)
\nonumber\\
&&\quad\quad\quad\quad\quad\quad+(\beta_M+\gamma_M)\cos(\phi-M\tau)\Bigr]\biggr\},
\ea
where $M$ and $N$ are (positive or negative) integers.  The system is
driven by two harmonics, each with its own resonant frequency. What
will happen? If the resonances are distinct enough, meaning they
affect motion in different parts of the phase space, they can coexist
peacefully. But supposing the resonances overlap - meaning there exist
initial conditions for which the motion in the phase space is
sensitive to both - what will the spin do? It does not know which
resonance to ``obey'', and hence its motion goes chaotic.  This is the
essence of the Chirikov criterion for the onset of wide-spread chaos
(Chirikov 1979; Lichtenberg \& Lieberman 1992).

\begin{figure}
\scalebox{0.4}{\includegraphics{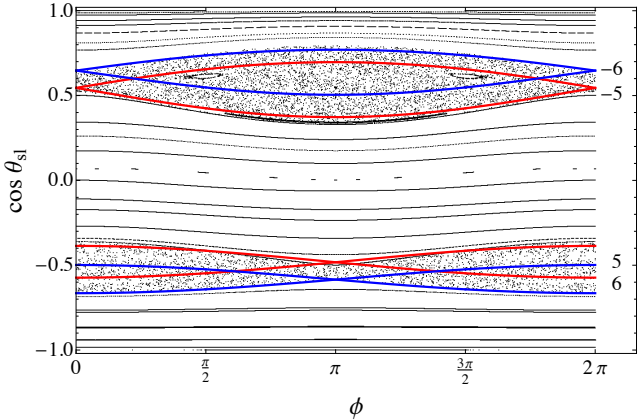}}
\caption{Surfaces of section for two pairs of resonances put together,
  with their respective analytically computed separatrices. Red:
  $N=5,\,M=-5$; blue: $N=6,\,M=-6$. The adiabaticity parameter is
  $\epsilon=0.1$.}
\label{4harm}
\end{figure}

\begin{figure*}
\includegraphics[width=\textwidth]{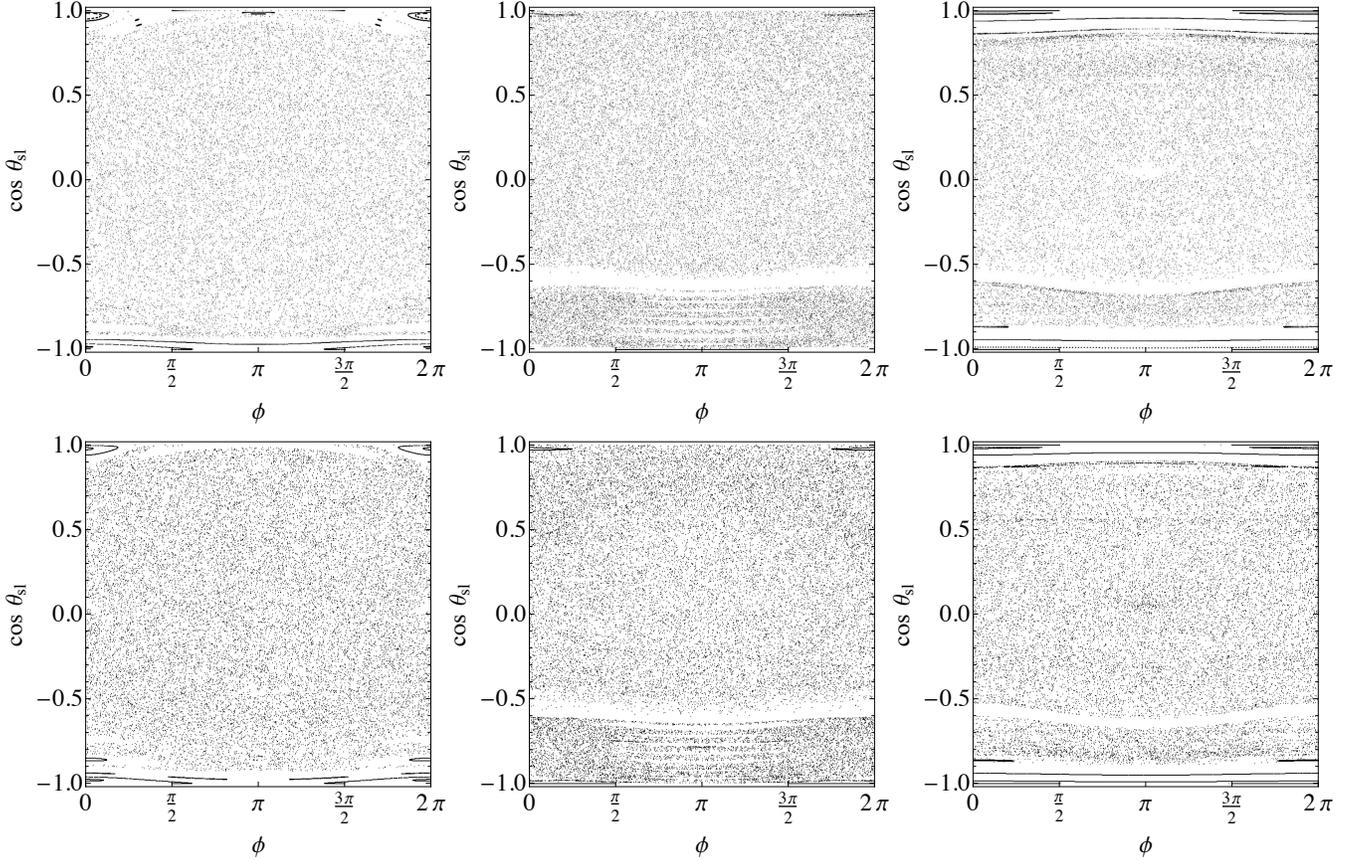}
\caption{Surfaces of section computed using the exact Hamiltonian
  (top panels) and using the approximate Hamiltonian with only the $\{-N_{\rm
    max},N_{\rm max}\}$ Fourier harmonics included in the forcing
  function (bottom panels). The panels from left to right correspond to
  $\epsilon=0.1$, $0.05$, $0.02$. Note the agreement between top and
  bottom panels becomes better with smaller values of $\epsilon$.}
\label{fullvnmax}
\end{figure*}

Figure \ref{cnoc} illustrates the onset of chaos due to overlapping
resonances.  Note that the separatrix of each of these resonances is
time-independent only in its own frame of reference.
Thus, to visualize the combined effect of both resonances
and be able to interpret them using resonance overlaps,
we construct surfaces of section. Specifically, we record $p$ and $\phi$ only
once per eccentricity cycle at $ \tau=0,2\pi,4\pi,\cdots$, because in this
case we have $H(\bar{\phi}) = H(\phi)$ for any harmonic. This enables
us to overlay analytic calculations of the separatrices on top of the
surface of section in a meaningful way. 
By doing this, we can say that Figure \ref{cnoc} indeed demonstrates that,
approximately, given two resonances $N$ and $M$ separated by a
distance $\Delta p$, chaotic evolution of $p=\cos\thetaSL$ is induced when
\be 
\Delta p \lo \frac{1}{2}\left(w_N+w_M\right). 
\ee
When this occurs, the region of chaotic evolution approximately spans the
areas of both separatrices. 

Figure \ref{4harm} shows an example when four resonances are included
in the Hamiltonian. In practice, a particular resonance likely only overlaps 
with the resonance nearest to it. Thus it is possible to observe features such as
those depicted in Figure \ref{4harm}: multiple isolated regions of chaos
separated by a large domain of periodic space.

\section{Application to the Full Problem of Lidov-Kozai driven Spin Precession}

We now examine the full problem of stellar spin dynamics driven by a
planet undergoing LK cycles, with the Hamiltonian given by Eq.~(\ref{kozaiham}). 
If the chaotic behaviour of this full
system is indeed determined by resonances and their overlaps, and, 
as discussed in Section 3.3, there exists a maximum resonance order 
$N_{\rm max}$, we expect that approximating this full system with one
consisting only of all harmonics with $|N|<N_{\rm max}$ should reproduce 
the key features of the system. Thus we consider the approximate Hamiltonian
\ba
&& H'_{\rm app}\simeq \frac{\bar{\alpha}}{n_e}\biggl[-\frac{1}{2} p^2 
+\epsilon\,\psi_0\, p \nonumber\\
&&\qquad -\frac{\epsilon}{2}\sqrt{1-p^2}\!\!\sum_{N=-N_{\rm max}}^{N_{\rm max}}
\!\!\!\left(\beta_N+\gamma_N\right)\cos(\phi-N\tau)\biggr].
\label{fourierham}\ea
We evolve equations of motion obtained from both Eq.~(\ref{kozaiham})
and Eq.~(\ref{fourierham}). Figure \ref{fullvnmax} compares the
resulting surfaces of section for several values of $\epsilon$.
It is apparent that taking only the innermost $2N_{\rm max}+1$ harmonics 
in the perturbing functions adequately reproduces the behavior of the full
system, with better agreement for smaller $\epsilon$.
 
 \begin{figure}
\scalebox{0.49}{\includegraphics{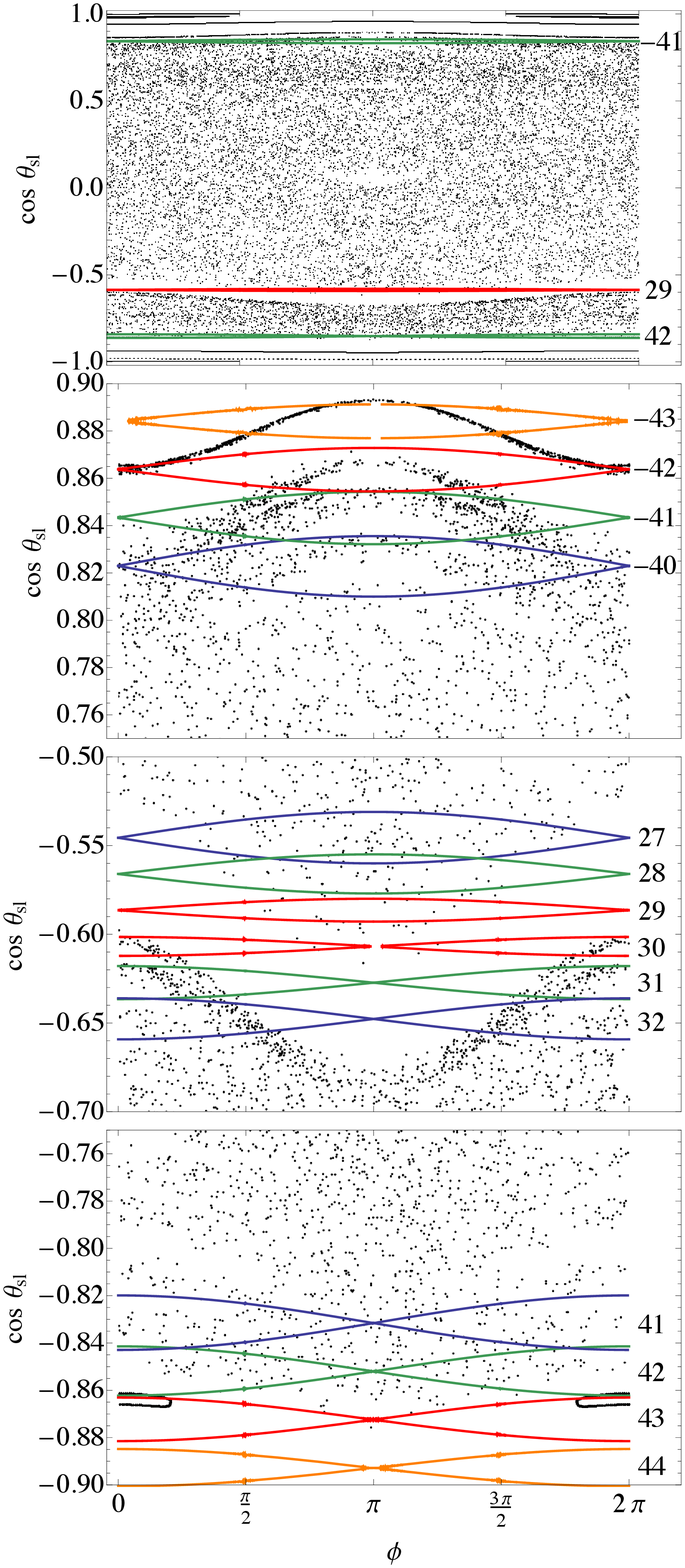}}
\caption{Demonstration of how overlapping resonances can explain several
  features of the $\epsilon=0.02$ surface of section shown in the
  right panels of Fig.~\ref{fullvnmax}. {\it Top panel}~: The entire surface
  of section, with the separatrices for the N=$42$ (bottom green),
  $29$ (red), and $-41$ (top green) resonances overlaid. {\it Second
    panel}~: Zoom-in on the top portion of the surface of section; the
  separatrices for resonances with $N=-40$ (blue), $-41$ (green),
  $-42$ (red), and $-43$ (orange) are overlaid. {\it Third panel}~:
  Zoom-in on the gap located at $p\approx -0.6$; from top to bottom,
  the separatrices for the $N=27$ (blue), $28$ (green), $29$ (red), $30$
  (red), $31$ (green), and $32$ (blue) resonances are overlaid. 
  {\it Bottom panel}~: Zoom-in on the bottom portion of the surface of
  section; the separatrices for $N=41$ (blue), $42$ (green), $43$ (red),
  and $44$ (orange) are overlaid. }
\label{resoverlap}
\end{figure}

We may now consider whether the overlap of these resonances can
explain the width of the chaotic region as a function of
$\epsilon$. Figure \ref{resoverlap} shows that this is indeed the case. 
Given a value of $\epsilon$, there exists a positive ``outermost'' resonance
$N=N^+_{\rm out}$ ($>0$) which overlaps with the ``previous'' resonance
($N^+_{\rm out}-1$) but not with the ``next'' one ($N^+_{\rm out}+1$). 
Since the separation (in $p$) of two neighboring resonances 
is $\Delta p\simeq 1/N_{\rm max}$ [see Eq.~(\ref{eq:pn})],
this ``outermost'' resonance is determined by the conditions
\be
{1\over 2}\left(w_{N^+_{\rm out}}+w_{N^+_{\rm out}-1}\right)>{1\over N_{\rm max}},
\ee
and
\be
{1\over 2}\left(w_{N^+_{\rm out}}+w_{N^+_{\rm out}+1}\right)<{1\over N_{\rm max}}.
\ee
Likewise, there exists a negative ``outermost'' resonance 
$N^-_{\rm out}$ ($<0$) which is the last to overlap with the ``previous'' one
($N^-_{\rm out}+1$). The locations of these two ``outermost''
resonances, as determined by the resonant momenta 
$p^\pm_{\rm N,out} \simeq -N^\pm_{\rm out}/N_{\rm max}$, bound
the chaotic region in the $p$-space
\footnote{Note that for sufficiently large $|N|$, the width of the resonance is
small [see Eq.~(\ref{appwidth})]. So the outer edge of the separatrix of the
outermost resonance is close to its center.}.

As $\epsilon$ is varied, $N^\pm_{\rm out}$ and $p^\pm_{\rm N,out}$
vary as well. Thus we may analytically compute the extent (i.e. the outermost
boundaries in the $p$-space) of wide-spread chaos as a function of $\epsilon$.
The result is shown in Figure \ref{chaoswidths}
(note that winthin the chaotic zone in $p$-space, there can still exist
periodic islands; see below).

\begin{figure*}
\includegraphics[width=\textwidth]{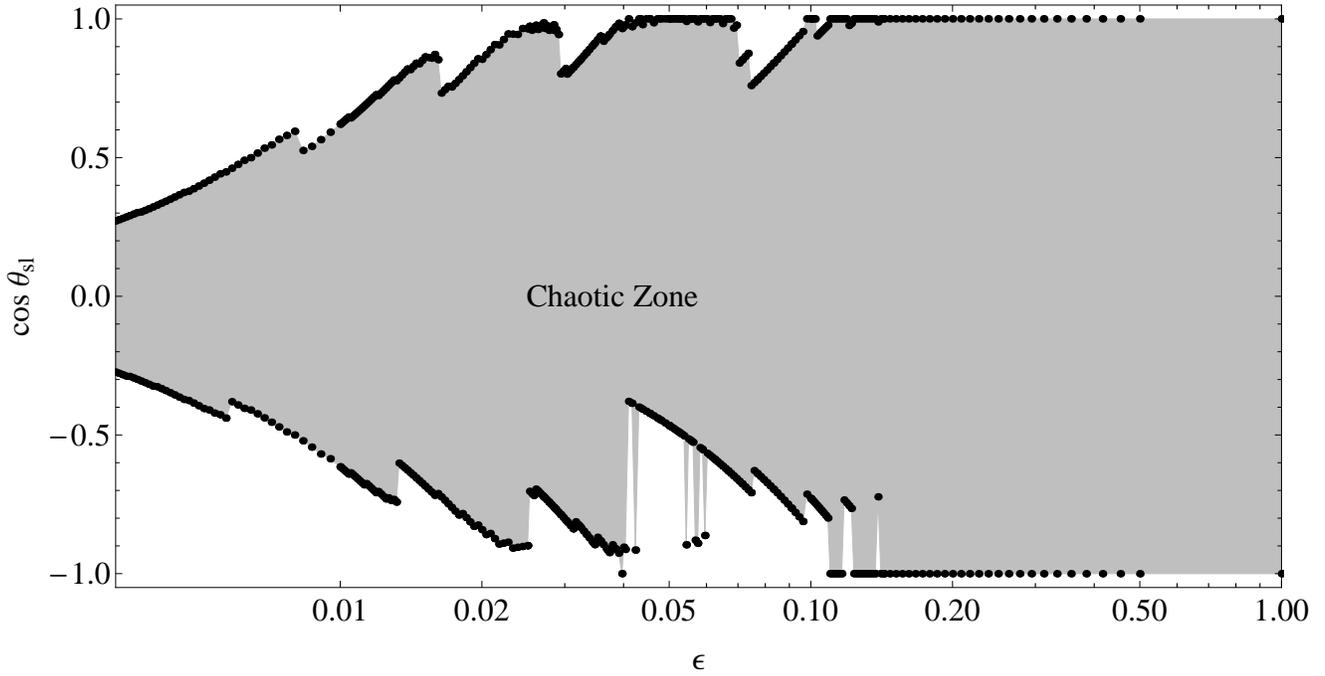}
\caption{Outermost boundaries of the chaotic region as a function of $\epsilon$,
  calculated by determining the outermost resonance $N^\pm_{\rm out}$
  which still overlaps with the previous one. The non-monotonic nature
  of the width of the chaotic region is due to the non-monotonic behavior of
  the Fourier coefficients of the resonant forcing terms
  ($\beta_N+\gamma_N$; see Fig.~\ref{coeffs}). 
  Note that while the spin evolution is strictly non-chaotic 
  outside the chaotic zone (the shaded region), there could be periodic 
  windows inside the chaotic zone.}
\label{chaoswidths}
\end{figure*}

\begin{figure*}
\includegraphics[width=\textwidth]{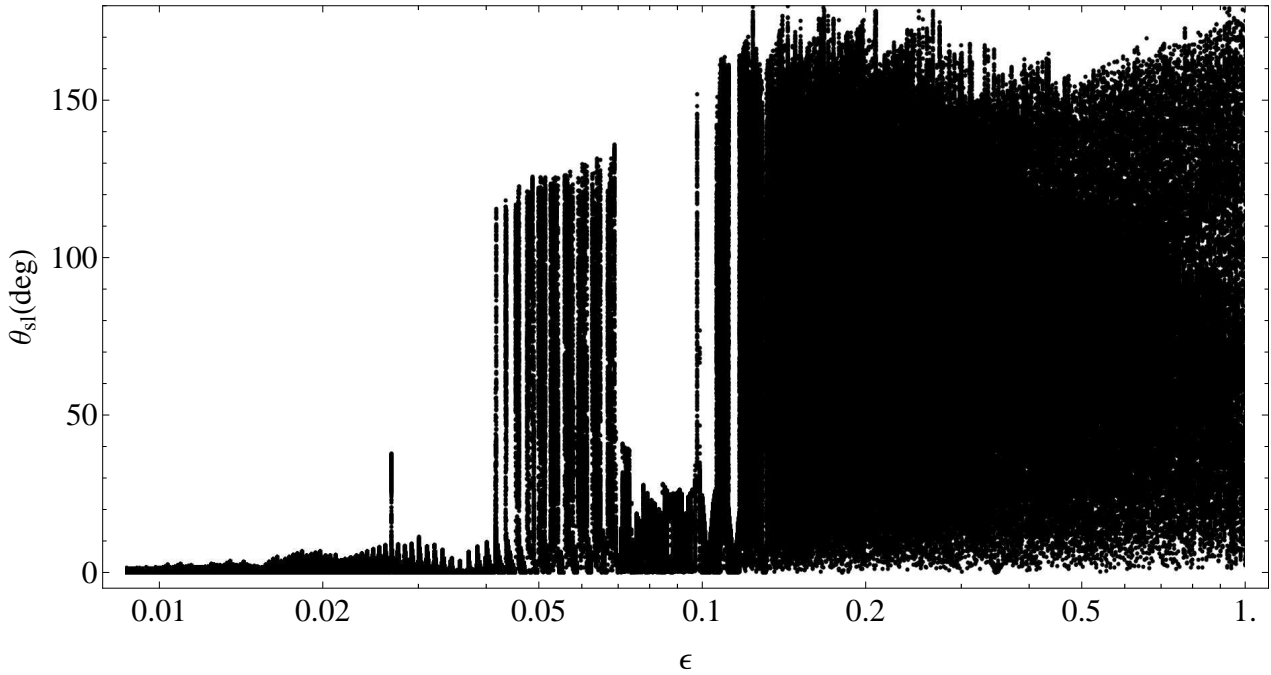}
\caption{``Bifurcation'' diagram of spin-orbit misalignment angle
  versus the adiabaticity parameter $\epsilon$. For each $\epsilon$, 
  we evolve the equations of motion starting with $\thetaSL=0$, for
  $\sim\!500$ LK orbital eccentricity cycles, and record $\thetaSL$ every time the
  eccentricity reaches a minimum. This diagram is similar to Fig.~1, except that all short-range force
effects and the back-reaction of the stellar spin on the orbit are
turned off.  }
\label{aligned}
\end{figure*}

Figure \ref{resoverlap} brings to light another interesting
feature of this dynamical system: the existence of narrow regions of
non-chaotic behavior, spanning the entire $\{0,2\pi\}$ range in the
$\phi$ coordinate and thus effectively splitting the phase space into
chaotic regions that cannot communicate with each other. This
feature arises from the strongly nonlinear variation of 
the Fourier coefficient ($\beta_N+\gamma_N$), and therefore the widths, 
of the various resonances involved: resonances that are very narrow are
isolated from the surrounding ones, and quasiperiodic behavior becomes
possible in their vicinity. For example, from Figure \ref{widths} we
see that for $\epsilon=0.05$, the resonances of order $N=11$ and $12$ are
particularly narrow, and indeed they are the ones that cause the
narrow band in the middle panels of Fig.~\ref{fullvnmax}. Likewise,
as demonstrated in the third panel
of Fig.~\ref{resoverlap},
for $\epsilon=0.02$, the resonances $N=29$ and $30$ are isolated from the
rest and result in a band of quasi-periodicity.

We now focus on systems which start out with aligned stellar spin and
planetary angular momentum axes (i.e. $\cos\thetaSL=1$) -- such
systems are very relevant in the standard picture where planets form
in protoplanetary disks aligned with the central stars.  Two questions
are of interest: first, given a specific value of $\epsilon$, will
such an initially aligned state experience chaotic or quasiperiodic
evolution, and second, if the evolution {\it is} chaotic, how much of
the available phase space will it span, i.e. how much will
$\cos\thetaSL$ vary? (A third question may also be asked - what
happens if $\epsilon$ slowly evolves as a function of time, as it
might in a physical system due to tidal dissipation? We address this
issue in Section 7 below).

To address these questions, we numerically construct a ``bifurcation
diagram'' (Fig.~\ref{aligned}), using the equations of motion of the
full Hamiltonian (Eq.~\ref{kozaiham}). 
For each value of $\epsilon$ we compute the spin evolution
trajectory starting from the initial condition $\cos\thetaSL=1$.
We record on the $y$-axis the spin-orbit misalignment angle at every
eccentricity minimum (at $\tau=0,2\pi,4\pi$...). The result is,
effectively, a 1D surface of section, for a single initial
condition. We then repeat the calculation for a fine grid of
$\epsilon$ values. Figure \ref{aligned} shows the result.  Large
spread in $\thetaSL$ indicates chaotic behavior, while small spread
with well-defined edges indicates quasiperiodicity.  We see from
Fig.~\ref{aligned} that, in general, the spread of $\thetaSL$ as a
function of $\epsilon$ follows the trend analytically predicted in
Fig.~\ref{chaoswidths}. For example, 
Figure \ref{chaoswidths} shows that for $\epsilon\go 0.1$, the spin-orbit
misalignment of an initially aligned state will evolve chaotically; 
this is consistent with Fig.~\ref{aligned}, which shows that $\thetaSL$
undergoes large excusion for $\epsilon\go 0.1$.
Figure \ref{chaoswidths} also shows
that only for $\epsilon \lo 0.02$, the aligned initial state will not
evolve into the chaotic zone; this is also reflected in
Fig.~\ref{aligned}, where for $\epsilon \lo 0.02$ the spread in
$\thetaSL$ is confined to a narrow region around $\thetaSL=0$.

\begin{figure}
\scalebox{0.56}{\includegraphics{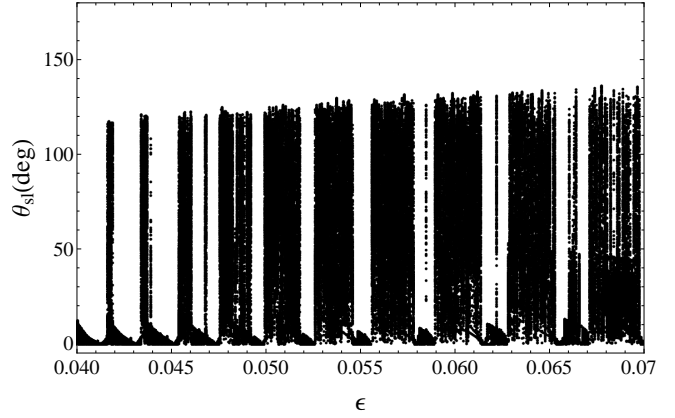}}
\caption{Zoom in on Fig.~\ref{aligned} in the region 
$0.04<\epsilon<0.07$. Here the spin evolution is mostly chaotic
(with large scatter in $\thetaSL$), with
periodic regions (with $\thetaSL$ close to zero)
appearing in the middle of the chaos.
The range of chaotic excursion is limited to be
less than $\sim 130^\circ$ due to a periodic island at
$\cos\thetaSL\sim -0.5$ caused by the narrow width of the $N=11$ and
$12$ resonances (see Fig.~\ref{widths} and middle panels of Fig.~\ref{fullvnmax}).}
\label{aligned3}
\end{figure}

\begin{figure}
\scalebox{0.56}{\includegraphics{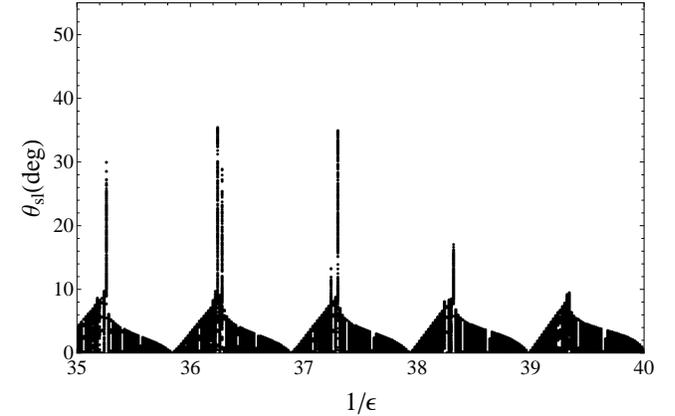}}
\caption{Zoom in on Fig.~\ref{aligned} in the region 
$0.025<\epsilon<0.0286$, plotted against $1/\epsilon$.
Here the spin evolution is mostly regular or periodic (with small scatter
in $\thetaSL$), but chaotic zones (with large scatter in $\thetaSL$) 
appear in the middle of the ``calm sea''. The occurrence of the chaotic
zones is approximately evenly spaced in $1/\epsilon$.}
\label{aligned2}
\end{figure}

However, the transition between adiabatic evolution and chaotic
evolution of stellar spin for an initially aligned state is fuzzy. As
seen in Fig.~\ref{aligned}, for $\epsilon$ between $\sim 0.02$ and
$\sim 0.1$, the regular (periodic) regions (with small spread in
$\thetaSL$) are interspersed with the chaotic zones (with large spread
in $\thetaSL$).  In particular, for $\epsilon \sim 0.04-0.07$, the
spin evolution is mostly chaotic but with somewhat regularly spaced
periodic regions -- ``periodic islands in an ocean of chaos'' (see
Fig.~\ref{aligned3}).  Toward smaller $\epsilon$, the periodic islands
expand and the chaotic regions shrink, so that for $\epsilon \lo 0.04$
the spin evolution becomes mostly periodic, with small finely tuned
chaotic domains that are shown in Fig.~\ref{aligned2} to be linearly
spaced in $1/\epsilon$ -- ``chaotic zones in a calm sea''.  To
illustrate how the theory of overlapping resonances can explain these
features,
Figure \ref{respassage} takes a closer look at the
resonances near $\cos\thetaSL=1$ for three closely spaced values of
$\epsilon$. Naturally, as discussed in Section 3.3, the resonance that
determines the evolutionary behavior of the initially-aligned system
is $N=-N_{\rm max}$, since it has $p_N\simeq 1$. As $\epsilon$ is
varied, the trajectory of the system falls either inside the
$N=-N_{\rm max}$ resonance, or outside it, or right on its
separatrix. The proximity of the $N=-N_{\rm max}$ resonance to the
$N=-N_{\rm max}+1$ resonance then determines the evolutionary
trajectory of the system. If the two resonances overlap strongly, then
all trajectories in the vicinity will be chaotic, but this is not the
case in Fig.~\ref{respassage}. Instead, for small values of
$\epsilon$, the $N=-N_{\rm max}$ separatrix appears to be close to,
but not quite touching, its neighbor. This, in principle, does not
completely preclude chaos, since the Chirikov criterion is, in fact,
too strict and chaos can still exist when two resonances are
sufficiently close to each other and the trajectories are close to one
of the separatrices (Chirikov 1979; Lichtenberg \& Lieberman 1992).
This is the case in Fig.~\ref{respassage}: the chaotic trajectory of
the middle panel falls right on the separatrix and effectively
``rides'' it out and onto the neighboring resonance.  Thus, the series
of peaks at small values of $\epsilon$ in Fig.~\ref{aligned} are due
to the varying proximity of the $N=-N_{\rm max}$ resonance to
$\cos\thetaSL=1$ and to its neighboring resonances.

\section{Adiabatic Resonance Advection}

\begin{figure}
\scalebox{0.53}{\includegraphics{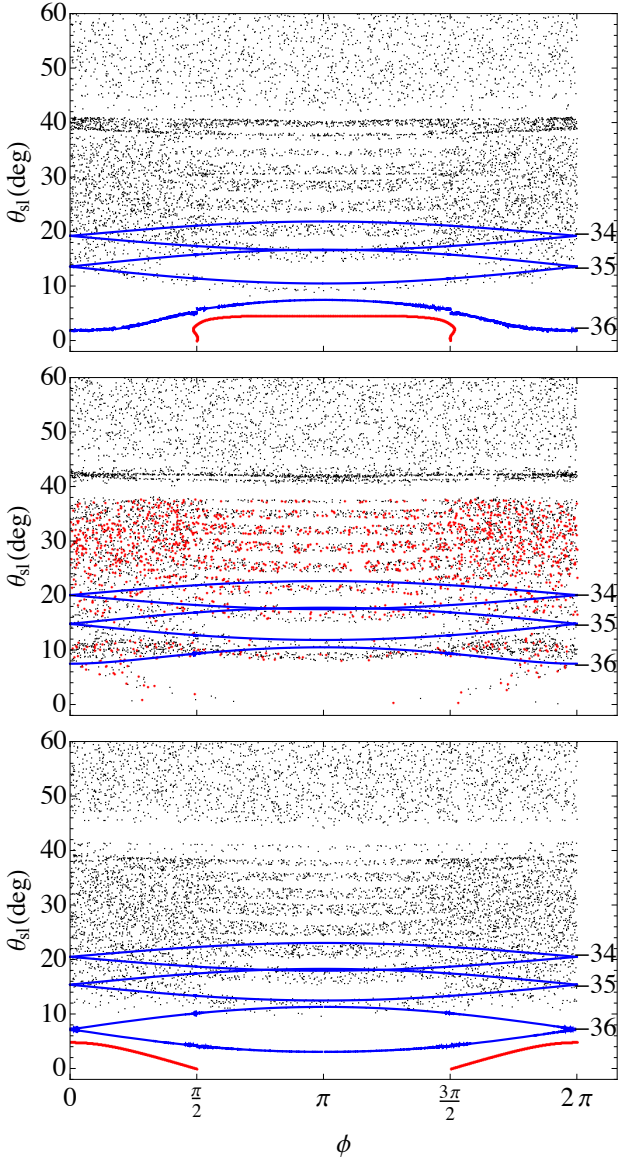}}
\caption{Demonstration of how 
the variation of resonances with $\epsilon$ 
leads to the peculiar oscillatory behavior seen in Fig.~\ref{aligned2}. 
The evolution of an initially aligned system 
is shown in red, and the analytical resonance separatrices are shown in
blue. {\it Top panel:}~ $1/\epsilon=37.1$; the initially aligned system
is trapped within the $N=-36$ resonance, which is sufficiently far
removed from the $N=-35$ resonance, so the trajectory is 
non-chaotic. {\it Middle panel:}~ $1/\epsilon=37.3$; the $N=-36$ and $N=-35$
resonances are close, so the trajectory becomes chaotic; $\thetaSL$ is
confined to $<35^\circ$ because of the gap which separates the two
chaotic zones. {\it Bottom panel:}~ $1/\epsilon=37.4$; the $N=-36$ resonance
has moved up sufficiently so that it no longer traps the initially aligned
system, and the trajectory is regular again.}
\label{respassage}
\end{figure}

\begin{figure}
\scalebox{0.66}{\includegraphics{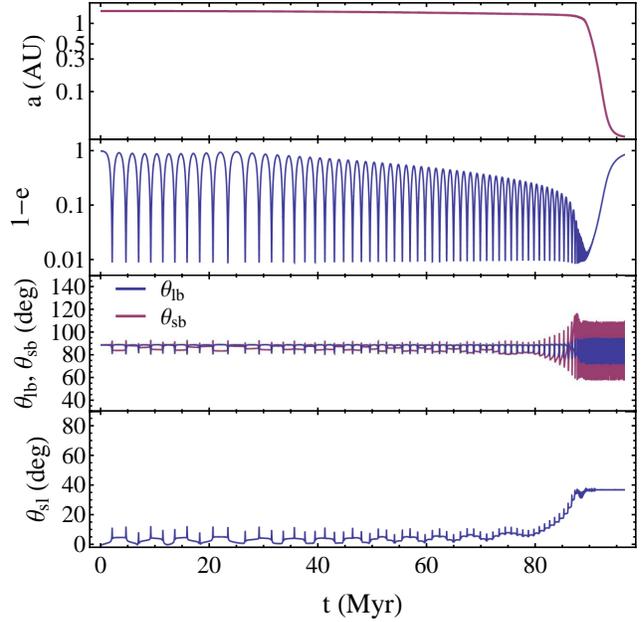}}
\caption{Sample time evolution demonstrating non-chaotic drift of an
  initially aligned system toward higher misalignment angles. The top
  panel shows the orbital semi-major axis, the second panel shows the
  eccentricity, the third panel shows the orbital inclination angle
  $\theta_{\rm lb}$ and the angle $\theta_{\rm sb}$ between $\hatS$
  and $\hatL_b$, and the bottom panel shows the spin-orbit
  misalignment angle $\thetaSL$.  The parameters are $M_p = 5M_J$,
  $\hat{\Omega}_\star = 0.05$, $a_0=1.5$AU, $a_b = 200$AU,
  $\theta^0_{\rm lb} = 89^\circ$, and we have included all short-range
  effects (cf. Fig. \ref{IATfull}). See SAL for details.}
\label{timeev}
\end{figure}

For a non-dissipative system, the adiabaticity parameter $\epsilon$ is
a constant. In the previous sections we have demonstrated that the
dynamical behavior of the stellar spin axis for different values of
$\epsilon$ can be understood using secular spin-orbit resonances.
Here we discuss the phenomenon of ``adiabatic resonance advection'',
and demonstrate the importance of resonances when dissipation is
introduced in our system.

As noted in Section 1, in the ``Lidov-Kozai + tide'' scenario for the
formation of hot Jupiters (Wu \& Murray 2003; Fabrycky \& Tremaine
2007; Correia et al.~2011; Naoz et al.~2012; Petrovich 2014; SAL),
tidal dissipation in the planet at periastron reduces the orbital
energy, and leads to gradual decrease in the orbital semi-major axis
and eccentricity. In this process, $\epsilon$ slowly decreases in
time. In SAL, we have considered various sample evolutionary tracks
and shown that the complex spin evolution can leave an imprint on the
final spin-orbit misalignment angle. A more systematic study will be
presented in a future paper (Anderson, Storch \& Lai 2015).

In Fig.~\ref{timeev}, we show a particular evolutionary track of our
system, obtained by integrating the full equations of motion
for the LK oscillations, including the effects of all short-range
forces (General Relativity, distortion of the planet due to rotation
and tide, and rotational bulge of the host star) and tidal dissipation
in the planet (see SAL for details). In this example, the adiabatic parameter
$\epsilon\simeq 0.17$ initially and decreases as the orbit decays.
So the spin evolution is always in the non-chaotic, adiabatic regime.
Interestingly, we see that as $a$ decreases, the initially aligned state
gradually drifts toward a higher misalignment angle in a well-ordered manner.

To explain this intriguing behavior, we consider a simplified version of the
problem, in which we gradually increase $\alpha_0$ (thereby decreasing
$\epsilon$) while keeping the forcing due to the planet
unchanged\footnote{This simplification implies that the ``shape''
functions [$\beta(\tau),~\gamma(\tau)$ and $\psi(\tau)$;
see Eqs.~(\ref{beta})-(\ref{psi})]
are unchanged as $\epsilon$ evolves. In real Lidov-Kozai
oscillations with tidal dissipation (depicted in Fig.~\ref{timeev}),
the range of eccentricity oscillations changes over time, with the
minimum eccentricity $e_{\rm min}$ gradually drifting from $e_0$
toward $e_{\rm max}$, thereby changing the shape functions.
To study this phenomenon quantitatively, this effect needs to be included.}. 
If the evolution of $\epsilon$ is sufficiently gradual,
then given an initial state there exists an
adiabatic invariant that is conserved as $\epsilon$ changes:
\be 
J=\oint p\, d\phi, 
\ee 
where the integration covers a complete cycle in the $\phi$-space.
This quantity is equivalent to the area enclosed by the trajectory in
phase space. Since, as discussed previously, the $N=-N_{\rm max}$
resonance is the one that most strongly influences the
initially-aligned system, we consider the single-resonance Hamiltonian
(Eq.~\ref{oneharmonic}) for this resonance. Since this Hamiltonian is
independent of time, it is conserved, i.e. $E=H(\phi_0,p_0)$ is a
constant so long as $\epsilon$ is constant. Conversely, a single value
of $E$ corresponds to a unique phase space trajectory $p(\phi;
E,\epsilon)$.  
It follows that the adiabatic invariant can be expressed as a function
of $E$ and $\epsilon$, i.e., $J = J(E,\epsilon)$.

\begin{figure}
\scalebox{0.49}{\includegraphics{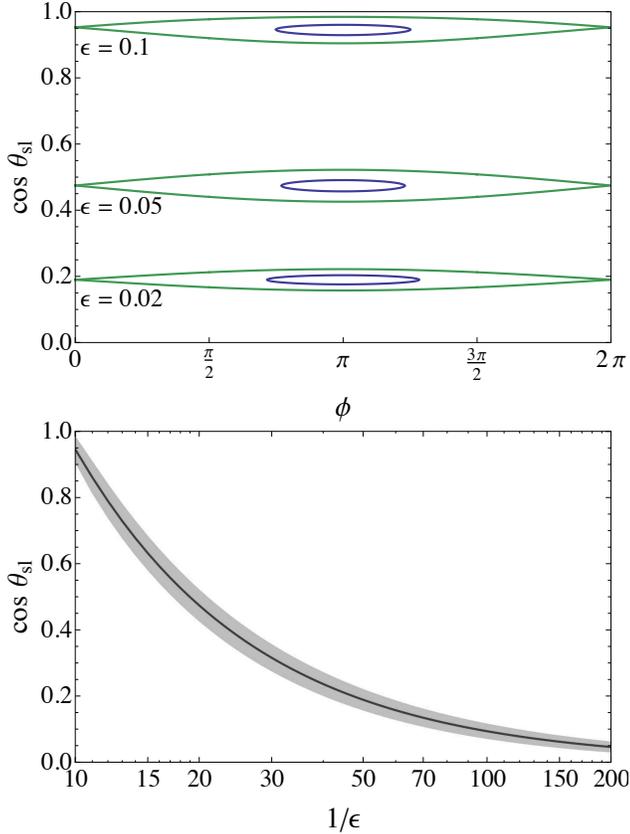}}
\caption{Proof of concept for 
``adiabatic resonance advection''.
{\it Top panel:}~ Sample spin evolution trajectories (constant-energy curves
in the $\cos\thetaSL$-$\phi$ phase space) for several values of
$\epsilon$.  The system initially has $\epsilon=0.1$ and is contained
within the $N=-9$ resonance with $p_N\simeq 1$.
As $\epsilon$ slowly decreases due to dissipation, the resonance center $p_N$
moves to smaller values, with the sample trajectory's area
remaining constant. 
{\it Bottom panel:}~ Location (solid black line) and width (grey area) of the
$N=-9$ resonance as a function of $1/\epsilon$, demonstrating that
$p_N$ moves toward $p=0$ and the resonance width (in $\cos\thetaSL$)
narrows with decreasing $\epsilon$. The sample trajectory trapped
inside the resonance must follow the resonance in accordance with
the principle of adiabatic invariance.}
\label{advection}
\end{figure}

As the system evolves ($\epsilon$ slowly changes), $J(E,\epsilon)$ is
kept constant, so $E$ must change.
These changes in $\epsilon$ and $E$ lead to changes in the phase space
trajectory.  For an initially circulating trajectory that spans the
entire $\{0,2\pi\}$ range in $\phi$, to conserve the area under the
curve the most that can happen is that an initially curved trajectory
must flatten, approaching $p={\rm const}$, where the constant is
roughly the average of $p$ over the initial trajectory. However, if a
trajectory is librating and only encloses a small area, it can be a
lot more mobile as $\epsilon$ evolves. As demonstrated in Fig.~\ref{respassage}, 
one way for the initially-aligned trajectory to be
librating is for it to be trapped inside the $-N_{\rm max}$
resonance. We also know that as $\epsilon$ decreases the resonance
must move toward $p=0$
[see Eqs.~(\ref{nmax})-(\ref{eq:pn})].  We therefore posit that it is
possible that the initially-aligned trajectory can be advected with
the resonance, and gradually taken to higher misalignment angles. A
proof of concept of this process is shown in Fig.~\ref{advection}. 
While a detailed study of this process 
(such as the condition for resonance trapping) is beyond the scope of this paper,
we note that it has many well-known parallels in
other physical systems, such as the trapping of mean-motion resonances
when multiple planets undergo convergent migration.

\section{Conclusion}

In this work we have continued our exploration of Lidov-Kozai driven
chaotic stellar spin evolution, initially discussed in Storch,
Anderson \& Lai (2014), by developing a theoretical explanation for
the onset of chaos in the ``adiabatic'' to ``trans-adiabatic'' regime
transition. The behaviour of the stellar spin evolution depends on the adiabaticity
parameter $\epsilon$ [see Eq.~(\ref{eq:epsilon1}) or
  (\ref{eq:epsilon})].  Using Hamiltonian perturbation theory, we have
identified a set of spin-orbit resonances [see Eq.~(\ref{rescond})]
that determine the dynamical behaviour of the system.  The resonance
condition is satisfied when the averaged spin precession frequency of the star
is an integer multiple of the Lidov-Kozai precession frequency of the
planet's orbit. We have shown that overlaps of these resonances lead
to the onset of chaos, and the degree of overlap determines how
wide-spread the chaos is in phase space.
Some key properties of the system include the facts that the width of an
individual resonance is a non-monotonic function of the resonance
order $N$ (see Fig.~\ref{widths}), and that there exists a maximum
order $N_{\rm max}$ [see Eq.~(\ref{nmax})] that influences the spin
dynamics. These properties lead to several unusual features (such as
``periodic islands in an ocean of chaos'') when the system transitions
(as $\epsilon$ decreases) from the fully chaotic regime to the fully
adiabatic regime (see Fig.~\ref{aligned}). Focusing on the 
systems with zero initial spin-orbit misalignment angle,
our theory fully predicts the region of chaotic spin evolution 
as a function of $\epsilon$ (see Fig.~\ref{chaoswidths}) and explains
the non-trivial features found in the numerical bifurcation diagram
(Fig.~\ref{aligned}). 
Finally, we use the spin-orbit resonance and the principle of
adiabatic invariance to explain the phenomenon of ``adiabatic resonance
advection'', in which the spin-orbit misalignment accumulates in a
slow, non-chaotic way as $\epsilon$ gradually decreases as a result of 
dissipation (see Section 7).

The system we considered in this paper is idealized. We have not
included the effects of short-range forces, such as periastron
advances due to General Relativity, and the planet's rotational bulge
and tidal distortion. We have also ignored the back-reaction torque
from the stellar quadrupole on the orbit. These simplifications have
allowed us to focus on the spin dynamics with ``pure'' orbital
Lidov-Kozai cycles. Finally, we have only briefly considered the
effects of tidal dissipation, using an idealized model in which
the ``shape'' of the Lidov-Kozai oscillations does not change as the
semi-major axis decays.  All of these effects will eventually need to be
included, if we hope to not only understand the origin of the chaotic
behavior but also make predictions for the observed spin-orbit
misalignment distributions in hot Jupiter systems. We begin to
systematically explore these issues numerically in a future paper
(Anderson, Storch \& Lai 2015).

\section*{Acknowledgments}

We thank Kassandra Anderson for useful discussion and continued
collaboration.  This work has been supported in part by NSF grant
AST-1211061, and NASA grants NNX14AG94G and NNX14AP31G.


\end{document}